# Distribution of neighborhood size in cities

Anand Sahasranaman[1,2,*] and Henrik Jeldtoft Jensen[2,3#]

[1]Division of Science, Division of Humanities and Social Science, Krea University, Andhra Pradesh 517646, India.

[2]Centre for Complexity Science, Dept of Mathematics, Imperial College London, London SW72AZ, UK.

[3]Institute of Innovative Research, Tokyo Institute of Technology, 4259, Nagatsuta-cho, Yokohama 226-8502, Japan

[*] Corresponding Author. Email: anand.sahasranaman@krea.edu.in

[#] Email: h.jensen@imperial.ac.uk

**Abstract:**

We study the distribution of neighborhoods across a set of 12 global cities and find that the distribution of neighborhood sizes follows exponential decay across all cities under consideration. We are able to analytically show that this exponential distribution of neighbourhood sizes is consistent with the observed Zipf's Law for city sizes. We attempt to explain the emergence of exponential decay in neighbourhood size using a model of neighborhood dynamics where migration into and movement within the city are mediated by wealth. We find that, as observed empirically, the model generates exponential decay in neighborhood size distributions for a range of parameter specifications. The use of a comparative wealth-based metric to assess the relative attractiveness of a neighborhood combined with a stringent affordability threshold in mediating movement within the city are found to be necessary conditions for the the emergence of the exponential distribution. While an analytical treatment is difficult due to the globally coupled dynamics, we use a simple two-neighbourhood system to illustrate the precise dynamics yielding equilibrium non-equal neighborhood size distributions.

## 1. Introduction

The distribution of city sizes in a country has been termed as an uncharacteristic regularity in economics [1, 2]. Across many national contexts, city sizes are found to be distributed according to a power law, specifically the rank-size distribution of city sizes is said to obey Zipf's Law [1, 2, 3, 4, 5, 6]. If we were to rank cities based on their population size, Zipf's Law posits that (Eqs. 1,2):

$$R_i = AP_i^{-\alpha} \qquad (1)$$

$$\ln R_i = \ln A - \alpha \ln P_i \qquad (2)$$

where $R$ and $P$ refer to the rank and population size, and $A$ is a constant, then the estimated coefficient $\alpha \approx 1$. A meta-study of 515 estimates from 29 studies of city size distribution from around the world finds the mean estimate of $\alpha$ to be 1.1 with two-thirds of the estimates ranging between 0.8 and 1.2 [5]. Another cross-country analysis covering 75 countries was found to yield an average exponent of 1.1 [6]. This regularity has sought to be explained using multiple theoretical models - in terms of the competing dynamics of new cities born at the rate of $\nu$ and existing cities growing at rate $\gamma$, yielding a power law with exponent $\alpha = \nu/\gamma$ [7]; using a stochastic growth model where migrants choose to form a new city with probability $\pi$ and enter an existing city otherwise, resulting in $\alpha \approx 1$ when $\pi \approx 0$ [8]; by assuming identical growth processes across city sizes (Gibrat's Law) resulting in Zipf's Law of city size distribution with $\alpha = 1$ [1]; and by using agent based models where each agent (firm) makes decisions on its location based on the location's demographics, yielding Zipf's Law under certain conditions [9].

In this work, we seek to fine-grain the scale of observation from the nation to the individual city, and study the distribution of neighborhood sizes across a city. To the best of our knowledge, this is a question that remains largely unexplored. Firstly, we attempt to statistically characterize the distributions of neighborhood sizes across cities. We then attempt to reconcile the findings on neighbourhood size distributions with the Zipfian distribution of city sizes. Finally, we attempt to theoretically model the potential mechanisms underlying characteristic neighbourhood size distributions.

## 2. Empirical neighborhood size distributions

The notion of a neighborhood (much like the boundary of a city) defies strict definition. While there may be broad agreement on neighborhoods being geographically localised communities within a city, the exact boundaries of neighborhoods in any given city remain open to debate. Despite this lack of specific definition, it is important to recognize that deeply local processes involving both local communities and local administration have led to the emergence of neighborhood areas and their corresponding governance structures. While these context-specific conceptualizations of neighborhoods may not be consistent across nations, they do offer us a mechanism to explore the distribution of neighborhoods (that have emerged out of lived local experience) in cities across the world.

In this work, we study the distribution of neighborhood size across 11 global cities - Cape Town, Rio de Janeiro, Mumbai, New York, Moscow, Shanghai, London, Buenos Aires, Berlin, Dhaka, and Toronto. These cities are spread across the world, ensuring diversity in historical contexts and socioeconomic outcomes. The notion of a neighborhood, as discussed earlier, is different across different cities - for instance, Mumbai's 97 wards each elect a Councillor, forming the level of government closest to the urban citizen, while New York City's Neighborhood Tabulation Areas (NTAs) were specifically created to be a summary level descriptor of the city's neighborhoods, offering a compromise between the broad strokes of the city's 59 districts and granularity of 2,168 census tracts, and Berlin's örststeiles are formally recognized localities for planning purposes, though not units of local government. Across the 12 cities under consideration (with average population 7.5 million), the average number of neighborhoods is 157, each with population 67,083. Appendix A provides the detailed data sources and neighborhood descriptions for all 12 cities. Table 1 presents a summary of neighborhood units used in this analysis.

| City | Neighborhood type | Neighborhood count | Average neighborhood population |
|---|---|---:|---:|
| Cape Town | Suburb, Township | 57 | 64,483 |
| Rio de Janeiro | Bairro | 159 | 40,103 |
| Mumbai | Ward | 97 | 128,272 |
| New York City | Neighborhood Tabulation Area (NTA) | 193 | 42,358 |
| Moscow | Raiyon | 123 | 86,042 |
| Shanghai | Township-level Division | 230 | 100,000 |
| London | Ward | 623 | 14,423 |
| Buenos Aires | Barrio | 48 | 57,826 |
| Berlin | Ortsteile | 96 | 35,034 |
| Dhaka | Thana | 41 | 161,711 |
| Toronto | Neighborhood | 172 | 13,800 |
| Singapore | Planning Area | 47 | 60,945 |

**Table 1**: *Neighborhood summary*

Unlike Zipf's Law, which appears to hold for distribution of city sizes, we do not find a consistent power law distribution of neighbourhoods. Instead, we find that the distribution of neighbourhoods across all cities under consideration is best described by exponential decay, and this holds across the entire range of neighbourhood sizes for each city (Fig. 1). What this essentially indicates is that, despite the large variations in neighbourhood count and neighbourhood sizes across cities (Table 1), the emergence of an exponential distribution of neighbourhoods appears to be a consistent phenomenon in cities around the world.

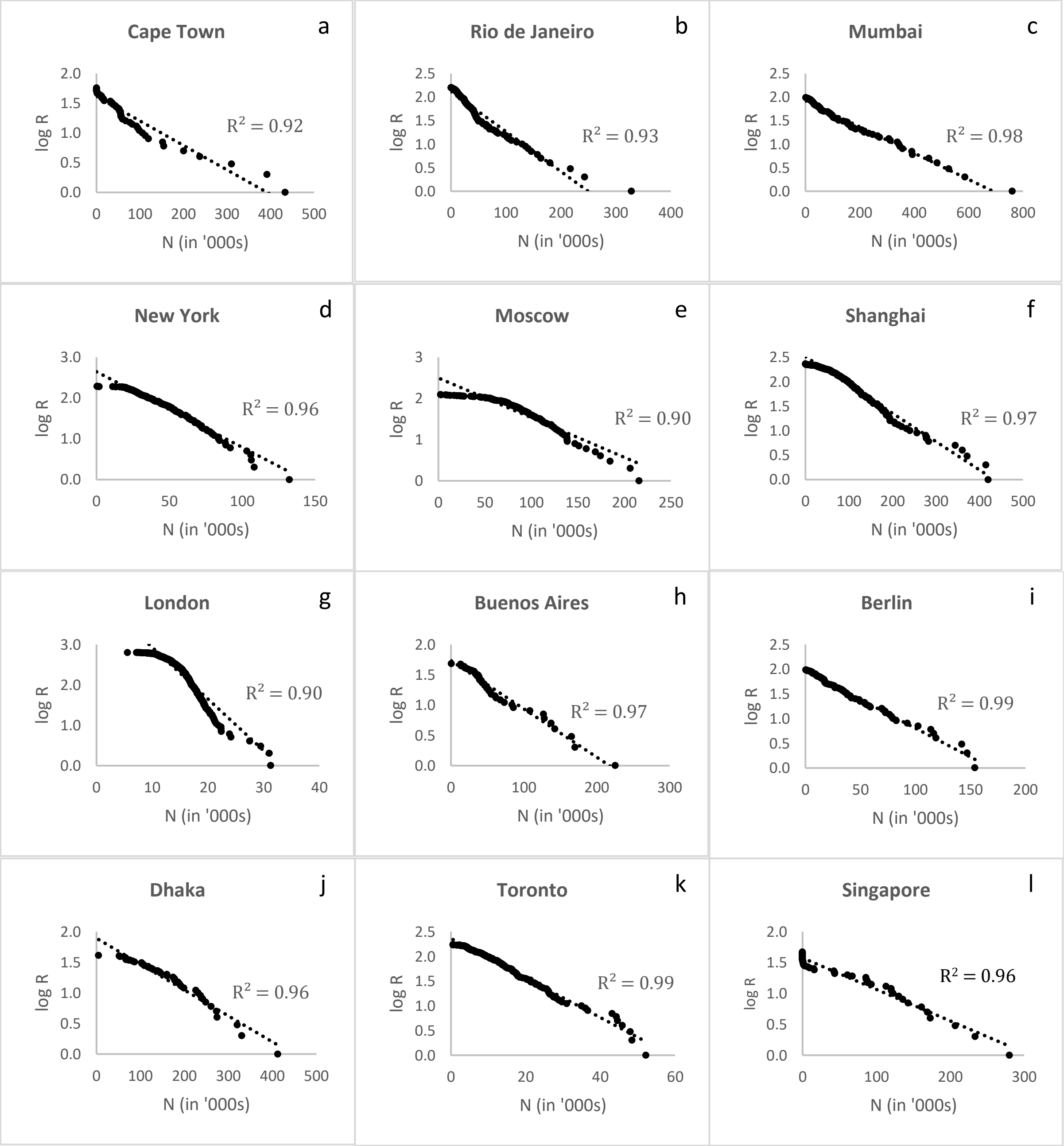

**Figure 1**. *log (Rank) v. Neighbourhood Size*. a: Cape Town. b: Rio de Janeiro. c: Mumbai. d: New York. e: Moscow. f: Shanghai. g: London. h: Buenos Aires. i: Berlin. j: Dhaka. k: Toronto. l: Singapore. Across all cities, the distribution of neighborhood size shows exponential decay. Black Dots: Actual neighbourhood size distributions. Dashed Black Line: Best fit line for exponential decay.

Given that all the evidence available suggests that the unequal distribution of neighbourhood sizes in cities is characterised by exponential decay, we now explore whether such neighbourhood distributions within cities can yield city-size distributions that follow Zipf's law.

**3. From exponential decay of neighbourhoods to Zipf's Law in cities**

We begin from our finding that each city has, approximately, an exponential distribution of neighbourhood sizes $(c)$. The probability of a neighbourhood with size $c = C$ is:

$$P\{c = C\} = \frac{1}{\lambda} e^{-C/\lambda} \quad (1)$$

Let $N$ denote the number of neighbourhoods in a city. The size of a city $(s)$ is the sum of its neighbourhood sizes:

$$s = \sum_{i=1}^{N} c_i \quad (2)$$

Given that the sum of $N$ exponentially distributed variables is a gamma distribution, the probability of a city of size $s = S$ is:

$$P\{s = S\} = \frac{\lambda^N S^{N-1} e^{-\lambda S}}{\Gamma(N)} \quad (3)$$

Now, consider a set of cities characterised by a specific set of values of $\lambda$ and $N$. Assume that the probability that a city picked at random has parameters $\lambda$ and $N$ is given by $P(N, \lambda)$. Therefore, the probability of a randomly selected city being of size $S$ is:

$$P\{s = S\} = \sum_N \sum_\lambda P(N, \lambda) \, \frac{\lambda^N S^{N-1} e^{-\lambda S}}{\Gamma(N)} \quad (4)$$

Assuming that $N$ and $\lambda$ are real positive numbers, we can replace the summations by integrals over the real axis. Therefore:

$$P\{s = S\} = \int_0^\infty dN \int_0^\infty d\lambda \;\; P(N, \lambda) \, \frac{\lambda^N S^{N-1} e^{-\lambda S}}{\Gamma(N)} \quad (5)$$

Replacing $\lambda$ with $x/S$, we get:

$$P\{s = S\} = \frac{1}{S^2} \int_0^\infty dN \int_0^\infty dx \;\; P(N, x/S) \, \frac{x^N e^{-x}}{\Gamma(N)} \quad (6)$$

The dependence on $S$ is only through the term $P(N, x/S)$, which may be a weak dependence, so we have the approximation:

$$P\{s = S\} \sim S^{-2} \quad (7)$$

This is equivalent to Zipf's Law for city size distributions. Indeed, the weak dependence on the $P(N, x/S)$ term possibly explains the variation of the power law exponent for city size distributions around the exact value of 1, as observed in large meta-studies for city systems around the world [6, 5]. Eqs. 1-7 therefore analytically demonstrate that the exponential decay observed in neighbourhood sizes is consistent with Zipfian distribution of city sizes.

## 4. Model for neighbourhood dynamics

This still leaves open the question of why exponential decay describes neighborhood size distributions. We build a model of neighborhood evolution based on individual choices that are determined by the context of the neighborhoods that individuals inhabit, and attempt to isolate the general mechanisms that result in the emergence of exponential decay in neighbourhood size distributions. Our modelling demonstrates that the observed exponential behaviour may be a consequence of individuals making decisions to move within the city based on an assessment of the relative wealth of their neighborhoods.

The model we present here is closely related to other metapopulation models we have built earlier to explore long-term economic status of urban neighbourhoods as well as the emergence of multiple segregations in cities [10, 11], and belongs in the long tradition of models going back to the Schelling segregation model [12]. We consider a city of $M$ neighborhoods with total population $P(0)$, where each neighborhood $i$ ($i \in 1,2, \ldots, M$) is initially composed of an equal number of agents, $P(0)/M$. Each agent is characterized by its wealth. The total wealth of neighborhood $i$ is the sum of the wealths of all agents in $i$ and denoted by $w_t(i)$.

Each iteration of the model comprises migration into the city and movement within the city. First, agents attempt to migrate into the city, and the population attempting entry into the city is defined as a fraction $r_{mig}$ of the city's extant population. However, the actual number of agents able to enter the city is determined by their individual wealths. If an incoming agent's wealth ($w_m$) is greater than the median wealth of a randomly chosen neighbourhood $j$ in the city ($w_j^{med}$), then the agent enters that neighbourhood with probability 1. If not, the agent migration into the city is stochastic:

$$p_{ent} = \begin{cases} 1, & if\ w_m \geq w_j^{med} \\ e^{-\beta_m (w_j^{med} - w_m)}, & otherwise \end{cases}, \tag{8}$$

where $\beta_m$ is the calibrating factor for $p_{ent}$. Progressively decreasing $\beta_m$ is reflective of increasing relaxation of the affordability condition. This could, for instance, be interpreted as public policy in the form of social housing or rental vouchers that enables households to move into neighbourhoods that they are otherwise unable to afford. The base case value of $\beta_m$ is chosen such that movement into neighbourhoods in contravention of the wealth threshold condition is difficult (but not impossible), and is potentially reflective of real-world cities.

Once all agents have attempted migration into the city for a given iteration (time $t$), then the dynamics of movement within the city begin. At any given time $t$, $P_t$ agents are randomly chosen to attempt movement within the city. The decision of a random agent (in neighbourhood $i$) to move out of its location is based on the neighbourhood's relative wealth. A random receiving neighbourhood $j$ is chosen, and the agent chooses to move with probability 1 if the median wealth of $j$ ($w_j^{med}$) is greater than or equal to median wealth of $i$ ($w_i^{med}$), and with probability 0 otherwise.

If the agent chooses to move, then the actual occurrence of the movement is mediated by its ability to afford the move. If agent wealth ($w_a$) is at least equal to the median wealth of $j$ ($w_j^{med}$), then the agent moves with probability 1; the move becomes probabilistic otherwise:

$$p_{move} = \begin{cases} 1, & if\ w_a \geq w_j^{med} \\ e^{-\beta_m (w_j^{med} - w_a)}, & otherwise \end{cases} \tag{9}$$

$\beta_m$, the same parameter used to calibrate migration into the city, also calibrates movement within the city.

Table 2 lists the parameter values for the model simulations. The parameters are meant to reflect the range of urban dynamics resulting in neighbourhood size distributions in real cities. We vary the rate of migration into the city as well as the correlation between wealth and status, with scenarios depicting both strong correlation between wealth and resident status and no correlation between wealth and resident status at all. Finally, we also study the sensitivity of outcomes to changes in the calibration parameter for migration and movement, $\beta_m$.

| **Parameter** | Base case | Varying $r_{mig}$ | Varying status-wealth correlation | Varying $\beta_m$ |
|---|---|---|---|---|
| Number of neighbourhoods, $M$ | 50 | 50 | 50 | 50 |
| Initial population of agents, $P(0)$ | 2500 | 2500 | 2500 | 2500 |
| Rate of population attempting entry per iteration, $r_{mig}$ | 0.005 | 0.01 | 0.005 | 0.005 |
| Agent wealth distributions | $N(10,1)$ for residents; $N(7,1)$ for migrants | $N(10,1)$ for residents; $N(7,1)$ for migrants | $N(10,1)$ for residents; $N(10,1)$ for migrants | $N(10,1)$ for residents; $N(7,1)$ for migrants |
| $\beta_m$ | 10 | 10 | 10 | 100; 5; 2; 1 |
| Number of iterations | 300 | 300 | 300 | 300 |

**Table 2**: *Model parameters*

## 5. Results and discussion:

We find that neighbourhood distribution appears to be well approximated by an exponential distribution across a range of parameter specifications. When we consider the base case scenario, we find exponential decay in neighbourhood size, which is in agreement with empirical observation (Fig. 1a). In the base case, $\beta_m = 10$, which indicates a non-zero probability of an agent ($a$) being able to move into neighbourhood $j$ in contravention of the wealth threshold of that neighbourhood ($w_j^{med}$) (Eq. 9). We also find that doubling the fraction of population of migrants trying to enter the city at each iteration, $r_{mig} = 0.01$, yields an approximately exponential distribution as in the base case where $r_{mig} = 0.005$ (Fig. 1b). The nature of emergent dynamics therefore appears robust to changes in migration rates and system size. Finally, in the base case, migrant wealths are, on average, lower than resident wealths; so we test model outcomes by removing the correlation between resident status and wealth by drawing both resident and migrant wealths from the same distribution. We find that, even in this case, an exponential distribution results (Fig. 1c).

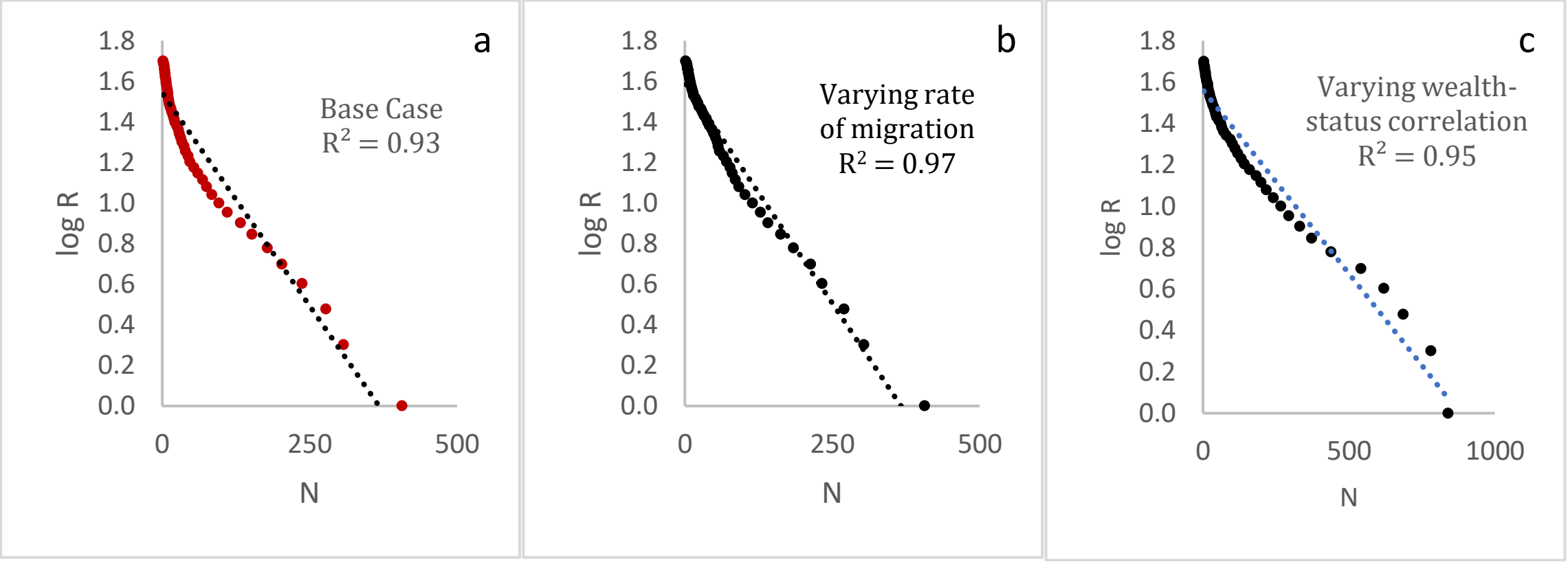

**Figure 2**. *Simulated plots of log (Rank) v. Neighbourhood Size (N)*. a: Base case. b: varying $r_{mig} = 0.01$, which is double the base case $r_{mig}$ of $0.005$. c: varying status-wealth correlation, ensuring no correlation between resident/migrant status and wealth (both drawn from the same wealth distribution), when compared to the base case when residents had higher wealth, on average. Across all scenarios, we find that the emergence of exponential neighbourhood size distributions is a robust result.

Overall, the emergence of an exponential distribution of neighbourhood sizes is consistent across a range of parameter specifications, but all these scenarios assume wealth as the basis for migration into and movement within the city. While these results indicate that wealth, both as the criterion for agent choice of neighbourhood and the mediator of agent movement based on affordability, is a sufficient condition for the emergence of exponential neighbourhood distributions, the question of whether it is a necessary condition remains open, and once we explore next.

We test the efficacy of the wealth threshold criterion in generating exponential neighbourhood size distributions: specifically, we explore how varying the stringency of the affordability condition, determined by $\beta_m$, impacts the emergence of neighbourhood sizes. We vary $\beta_m$ from $100$ to $1$ (taking values 100, 10, 5, 2, 1) (Table 2), and find that the emergence of the exponential distribution is consistent within a certain range of $\beta_m$, where only a low fraction amount of movement in contravention of neighbourhood thresholds is possible (Figs. 3a, 3b). Beyond this range, neighbourhood distributions are non-exponential (Figs. 3c, 3d, 3e). In order to quantify this range of $\beta_m$, we compute the threshold crossing ratio ($TCR$) as the ratio of number of times when an agent is able to successfully move into a neighbourhood $j$ despite its wealth ($w_a$) being lower than $w_j^{med}$ (i.e. $w_a - w_j^{med} < 0$) to the total number of such attempts to move in contravention of the wealth threshold over the time of the dynamics (Fig. 3f). For $\beta_m \geq 10$, we find that $TCR \leq \sim 0.07$, and the resulting distribution is best approximated by an exponential (Figs. 3a, 3b). However, as we progressively relax the wealth threshold condition for $\beta_m \leq 5$ ($TCR > \sim 0.12$), we find that there is greater condensation of population into smaller fractions of neighbourhoods, thus yielding closer approximations of potential power laws rather than exponentials (Figs. 3c, 3d, 3e, 3f). For instance, the top 10% of neighbourhoods (by population) account for ~74% of the population when $\beta_m = 1$, when compared to only ~48% when $\beta_m = 10$. This is because, at lower $\beta_m$, larger fractions of poorer agents are able to move neighbourhoods in contravention of the neighbourhood wealth condition, resulting in population aggregation in a few neighbourhoods. Overall, we find that while a small number of moves in contravention of wealth threshold conditions appears essential for an exponential distribution to emerge ($\beta_m = 100, TCR = \sim 0.01$), it is apparent that beyond a certain threshold ($\beta_m > 10, TCR > \sim 0.12$) the distribution of neighbourhood size does not result in exponential decay. This finding confirms that the existence of both a decision-making condition based on relative wealths of neighbourhoods ($w_i^{med}$ v. $w_j^{med}$), and an affordability condition mediating movement within the city ($\beta_m \geq 10, TCR < \sim 0.07$), are necessary conditions in the model to yield an exponential distribution of neighbourhood sizes.

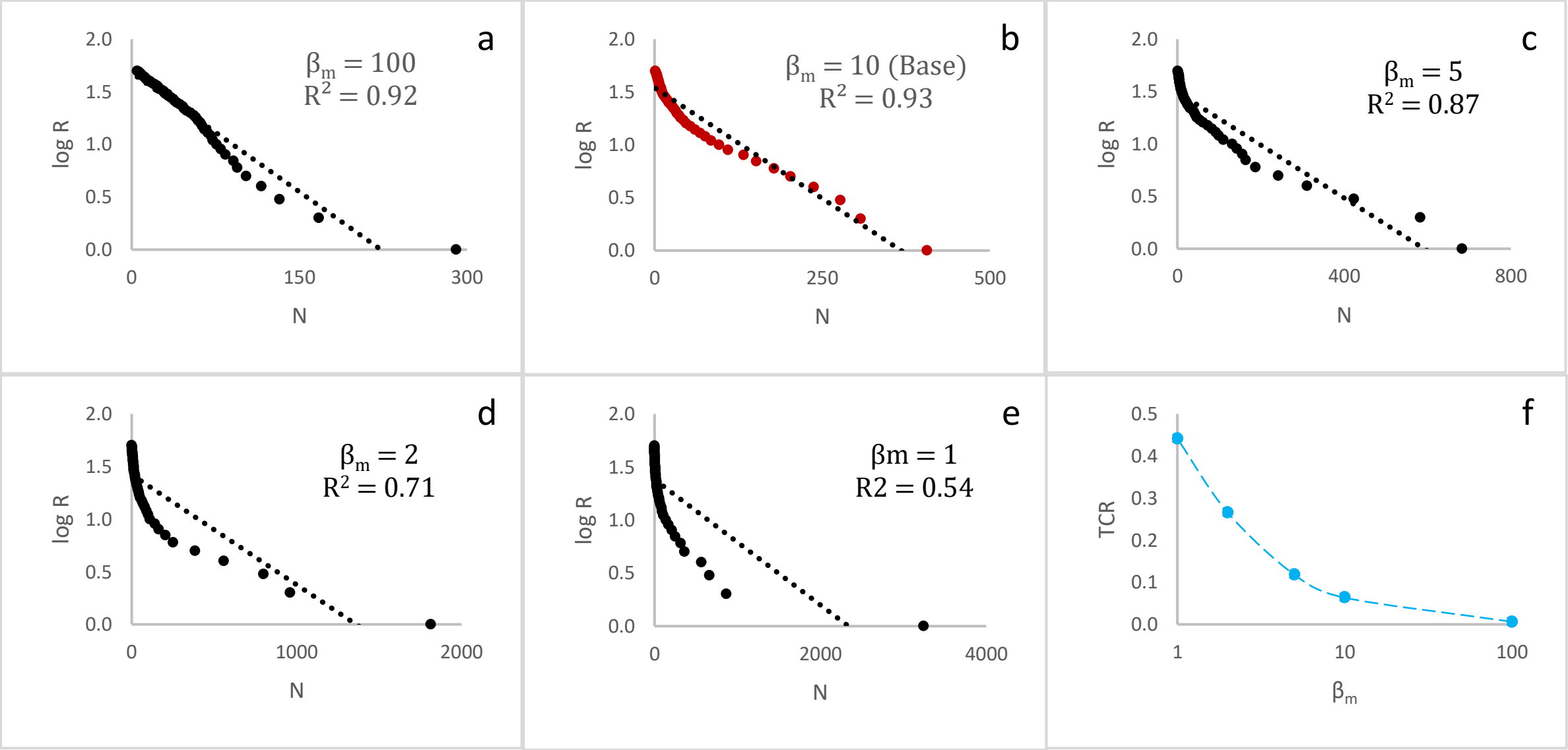

**Figure 3**. *Testing the necessity of wealth in generating exponential neighbourhood distributions*. a: log $R$ v. $N$ for $\beta_m = 100$. b: log $R$ v. $N$ for $\beta_m = 10$. c: log $R$ v. $N$ for $\beta_m = 5$. d: log $R$ v. $N$ for $\beta_m = 2$. e: log $R$ v. $N$ for $\beta_m = 1$. As $\beta_m$ decreases, the distribution of neighbourhood sizes moves away from exponential decay indicating that a reasonably stringent affordability condition is operational in cities, resulting in exponential distribution of neighbourhood populations. f: $TCR$ v. $\beta_m$. As $\beta_m$ decreases, $TCR$ increases.

The globally coupled nature of the dynamics under consideration makes exact analytical treatment very difficult. However, we can undertake a simplified analytical exploration of the dynamics in a highly structured two neighbourhood system. Consider a city system composed of two neighborhoods $H_1$ and $H_2$, populated by $N$ agents. There are two kinds of agents – $\frac{N}{2}$ agents are of type $A_1$ with wealth $w_1$; and the remaining $\frac{N}{2}$ agents are of type $A_2$ with wealth $w_2$, such that $w_1 > w_2$. Initially, the agents are equally distributed across both neighborhoods, each with $\frac{N}{2}$ agents. At every point in time, each agent decides whether it wants to move from its current location $i$ to $j$ based on the wealth comparison between $i$ and $j$: an agent moves only if $w_j^{med} \geq w_i^{med}$.

Let us first consider a scenario where each neighbourhood is populated by an equal number of $A_1$ and $A_2$ agents (Fig. 4a, initial). In this scenario, both neighbourhoods have the same median wealth, $w_{H_1}^{med} = w_{H_2}^{med}$. Therefore, all agents are satisfied with their current locations and the initial configuration is an equilibrium configuration (Fig. 4a, final).

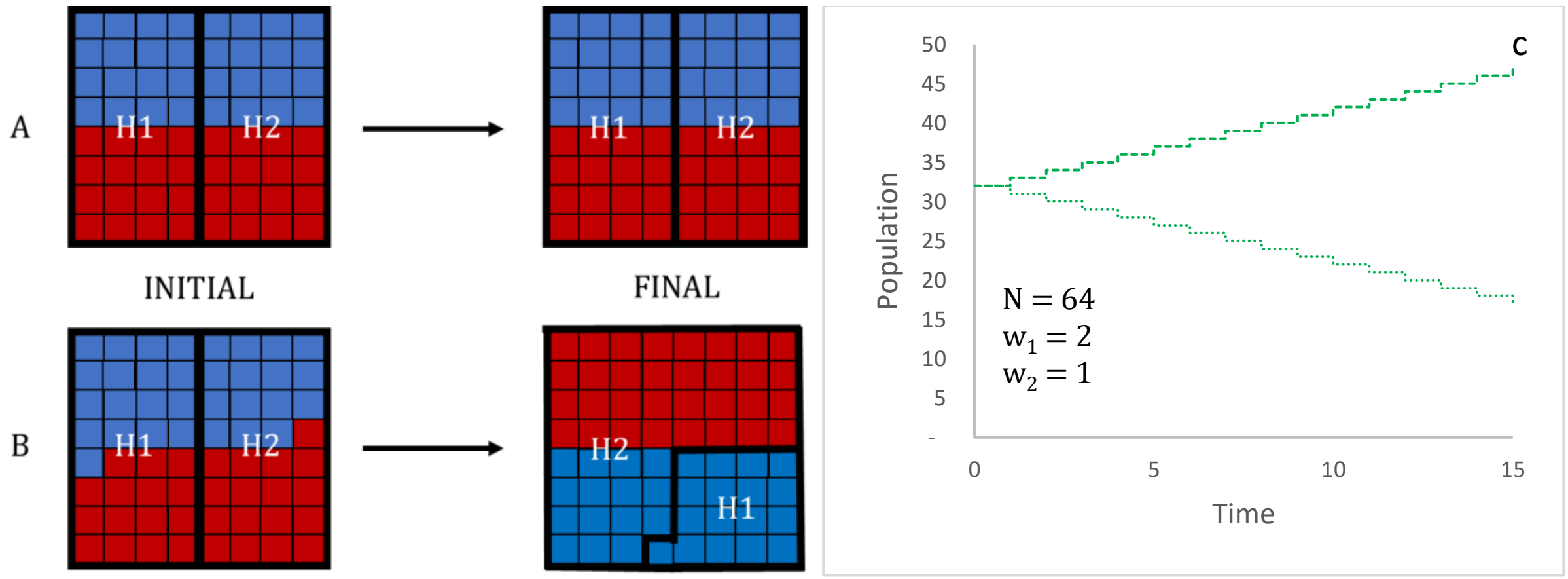

**Figure 4.** *Analytical description of dynamics in two-neighborhood system*. a: Initial state and Final equilibrium for a system ($N = 64, w_1 = 2, w_2 = 1$) where each neighborhood starts with an equal number of $A_1$ (red) and $A_2$ (blue) agents. Each square represents an individual agent and the thick black lines represent neighborhood boundaries. b: Initial and final configurations for a system which starts with an equal number of agents, but $H_1$ has one $A_2$ agent more than $H_2$ and $H_2$ has one $A_1$more agent than $H_1$. c: Evolution of total population over time in the two-neighborhood system. Evolution of total population in $H_1$ (dotted green) and population in $H_2$ (dashed green) shows that both neighbourhoods have population of 32 to begin with, but final population of $H_1$ is 17, while that of $H_2$ is 47.

Now, let us perform the slightest perturbation of the initial condition and swap one $A_1$ agent from $H_1$ with one $A_2$ agent from $H_2$ (Fig. 4b, initial). The populations of the two cells remains equal at $\frac{N}{2}$, but the fraction of $A_1$ agents in $H_1$ is $\frac{\frac{N}{4}-1}{\frac{N}{2}} < 0.5$ and in $H_2$ is $\frac{\frac{N}{4}+1}{\frac{N}{2}} > 0.5$. Therefore, the median wealths of $H_1$ and $H_2$ are: $w_{H_1}^{med} = w_2$ and $w_{H_2}^{med} = w_1$ respectively. Given this configuration, all agents in $H_2$ are satisfied with their current location, but all agents in $H_1$ are unsatisfied in terms of neighbourhood wealth comparison ($w_{H_2}^{med} > w_{H_1}^{med}$). Despite this dis-satisfaction with their current location, $A_2$ agents in $H_1$ are unable to move to $H_2$ because their wealths ($w_2$) are lower than the median wealth of $H_2$ ($w_{H_2}^{med} = w_1 > w_2$). Therefore, the final equilibrium in this two-neighbourhood system involves all $A_1$ agents in $H_1$ moving to $H_2$, because their wealths $w_1$ are equal to the median wealth of $H_2$, which remains at $w_{H_2}^{med} = w_1$ for the duration of the dynamics. This results in an unequal distribution of population across the two neighbourhoods, with $H_1$'s final population at $\frac{N}{4} + 1$ and $H_2$'s at $\frac{3N}{4} - 1$. Figure 4c illustrates the evolution of a two-neighborhood system for specific values of $N, w_1, w_2$ and its final equilibrium in terms of total population. What this simple analytical illustration demonstrates is that even the mildest perturbation in the initial condition in a highly structured two-neighborhood system with equal populations at the outset, results in an equilibrium with far from equal population sizes across neighborhoods.

## 6. Conclusion:

We study neighborhoods across a set of 12 global cities and find that the distribution of neighborhood sizes follows exponential decay across all cities under consideration. Given that city sizes are distributed according to Zipf's Law, we explore the question of whether exponential neighbourhood sizes are consistent with city size distributions. We are able to demonstrate analytically that city populations, composed of exponentially decaying neighbourhood sizes, can be distributed as a power law with exponent $\sim 1$.

In order to explore the emergence of the exponential distribution of neighbourhood size, we build a model of wealth-based neighbourhood dynamics. In this model, agents assess their satisfaction with their extant neighborhoods by using a simple wealth-based metric which compares their neighbourhood's median wealth with that of a randomly chosen neighbourhood in the city. If satisfied, agents stay back in their current neighborhood and if not, they attempt to move. Movement to another neighborhood is mediated by an affordability condition. Using this simple set up, we find that the dynamics yield exponential neighbourhood size distributions, in concordance with empirical observations. Most importantly, we find that using the wealth condition as the basis for decision making and movement within the city is essential for the emergence of exponential decay.

While a closed form analytical description eludes us due to the complexity of the dynamics, we construct a simple, highly structured two-neighbourhood system to illustrate the emergence of unequal neighbourhood sizes.

**Conflict of Interest:** On behalf of all authors, the corresponding author states that there is no conflict of interest.

**Competing Interests:** The authors declare that they have no competing interests.